\documentclass[a4paper, 10pt]{extarticle}
\usepackage{amsmath}
\usepackage{amssymb}
\usepackage{amsfonts}
\usepackage{graphicx}
\usepackage{subfigure}
\usepackage{authblk}
\usepackage[textwidth=16cm,textheight=20cm]{geometry}
\newtheorem{theorem}{Theorem}

\newtheorem{definition}[theorem]{Definition}

\title{Critical phenomena and singular solutions in non-stationary filtration of real gases}
\author[1,\thanks{\textit{E-mail: }\texttt{valentin.lychagin@uit.no}}]{Valentin Lychagin}
\author[1,\thanks{\textit{E-mail: }\texttt{mihail\underline{ }roop@mail.ru}}]{Mikhail Roop}
\affil[1]{V.A. Trapeznikov Institute of Control Sciences, Russian Academy of Sciences, 65 Profsoyuznaya Str., 117997 Moscow, Russia}

\begin{document}
\maketitle

\abstract{
In this paper, we study non-stationary filtration of real gases in porous media. Thermodynamic state of the medium is given by van der Waals state equations. Solutions for non-stationary filtration equation are obtained by means of finite dimensional dynamics. The analysis of phase transitions along the flow in case of isentropic and isenthalpic processes is presented as well as singular properties of solutions obtained are discussed. Domains in the jet space where the dynamics found is an attractor are shown.
}

\section{Introduction}
One-dimensional flows of one-component gas through porous media are described by the following system of differential equations (see, for example,~\cite{Lib,Mus,Sch}):
\begin{itemize}
\item the Darcy law
\begin{equation}
\label{darcy}
u=-\frac{k}{\mu}p_{x},
\end{equation}
\item the continuity equation
\begin{equation}
\label{cont}
\rho_{t}+(\rho u)_{x}=0,
\end{equation}
\end{itemize}
where $u(t,x)$ is the velocity of the gas, $p(t,x)$ is the pressure, $\rho(t,x)$ is the density, $k$ and $\mu$ are the permeability coefficient and viscosity respectively. Equation~(\ref{darcy}) corresponds to the momentum conservation law, equation~(\ref{cont}) is responsible for the conservation of mass. In addition to~(\ref{darcy})-(\ref{cont}) we assume that the medium is involved in one of two processes, isentropic or isenthalpic. In the first case the specific entropy $\sigma(t,x)$ is assumed to be constant, i.e. $\sigma(t,x)=\sigma_{0}$, while in the second one the specific enthalpy $\eta=e+p\rho^{-1}$, where $e$ is the specific energy, is assumed to be constant, $\eta(t,x)=\eta_{0}$. In both cases, we need additional relations for thermodynamic variables to make system~(\ref{darcy})-(\ref{cont}) complete. To this end, we use van der Waals equations of state. The van der Waals model allows to investigate such critical phenomena as phase transitions, and considering it together with equations describing dynamics one can analyze how phase transitions occur along the flow of the gas, which is the main goal of this paper.

Filtration processes with phase transitions were studied an a few works, for instance in~\cite{MZT,Kach}, where numerical methods were applied to the system of filtration equations. Some invariant solutions and analysis of admissible symmetries of filtration equations for various media are presented in~\cite{DLTfiltr}. Phase transitions in stationary filtration of real gases are studied in~\cite{LychGSA,LRljm,LRjgp,LRgsa}. The similar analysis for Euler and Navier-Stokes flows is presented in~\cite{LRamp}. In the present work, we use a method of finite dimensional dynamics~\cite{KrugLych,LychLych,AKL} to find exact solutions for non-stationary system~(\ref{darcy})-(\ref{cont}) for two types of processes --- isentropic and isenthalpic.
\section{van der Waals gases}
In this section, we briefly describe thermodynamic states geometrically (see~\cite{Lych,Mrug,Rup} for more details) and discuss phase transitions for the van der Waals model.

Let us consider the contact space $(\mathbb{R}^{5},\theta)$ with coordinates $(\sigma,e,v,p,T)$, where $v=\rho^{-1}$ is the specific volume and $T$ is the temperature, and with contact structure given by the differential 1-form
\begin{equation*}
\theta=d\sigma-(pT^{-1})dv-T^{-1}de.
\end{equation*}
By a thermodynamic state we mean a 2-dimensional submanifold $L\subset(\mathbb{R}^{5},\theta)$, such that $\theta|_{L}=0$. The last means that $L$ is a Legendrian manifold, on which the first law of thermodynamics holds. If one has $\sigma=\sigma(e,v)$ on $L$, then the Legendrian manifold $L$ is defined by the following relations:
\begin{equation*}
\sigma=\sigma(e,v),\quad T=\frac{1}{\sigma_{e}},\quad p=\frac{\sigma_{v}}{\sigma_{e}}.
\end{equation*}

The Legendrian manifold $L$ is also equipped with the differential quadratic form~\cite{Lych}
\begin{equation*}
\kappa=d(pT^{-1})\cdot dv+d(T^{-1})\cdot de.
\end{equation*}
The domains on $L$ where this form is negative are called \textit{applicable phases}. A jump from one applicable point $a_{1}=(\sigma_{1},e_{1},v_{1},p,T)\in L$ to another $a_{2}=(\sigma_{2},e_{2},v_{2},p,T)\in L$ governed by the intensives $(p,T)$ and the specific Gibbs potential $\gamma=e-T\sigma+pv$ conservation law is called \textit{phase transition}.

For van der Waals gases the Legendrian manifold $L$ is given by
\begin{equation}
\label{legvdw}
p=\frac{8T}{3v-1}-\frac{3}{v^{2}},\qquad e=\frac{4n}{3}T-\frac{3}{v},\quad \sigma=R\ln \left( T^{4n/3}\left(3v-1\right)^{8/3} \right),
\end{equation}
where $R$ is the universal gas constant, $n$ is the degree of freedom. Note that equations~(\ref{darcy})-(\ref{cont}) together with either isentropicity or insenthalpicity condition and~(\ref{legvdw}) become a complete system.

\begin{figure}[h!]
\centering
\includegraphics[scale=.3]{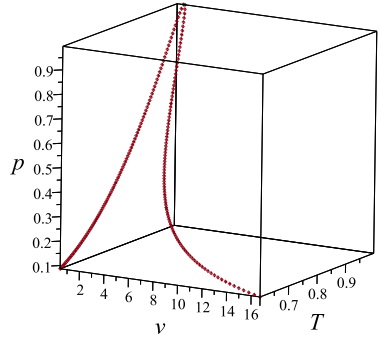}
\caption{Coexistence curve for van der Waals gases.}
\label{coexV}       
\end{figure}

The differential quadratic form is
\begin{equation*}
\kappa =-\frac{Rn}{2}\frac{dT^{2}}{T^{2}}-\frac{9R(4Tv^{3}-9v^{2}+6v-1)}{4Tv^{3}(3v-1)^{2}}dv^{2}.
\end{equation*}
We can see that $\kappa$ changes its sign, therefore there are phase transitions. Using the condition $\gamma(a_{1})=\gamma(a_{2})$ one can obtain equations for the coexistence curve, i.e. curve where phase transition occurs~\cite{LRljm}:
\begin{equation*}
\frac{p}{RT}=\frac{3}{3v_{1,2}-1}-\frac{9}{8v_{1,2}^{2}T},\quad 3(v_{1}-v_{2})=(3v_{2}-1)(3v_{1}-1)\ln\left(\frac{3v_{1}-1}{3v_{2}-1}\right).
\end{equation*}
The coexistence curve for van der Waals gases is shown in figure~\ref{coexV}.

By a thermodynamic process we shall mean a contact transformation $\Phi\colon\mathbb{R}^{5}\to\mathbb{R}^{5}$ preserving the Legendrian manifold $L$. From infinitesimal point of view, such a transformation is generated by a contact vector field $X$ tangent to $L$. Integral curve $l\subset L$ of $X$ is exactly what we call a thermodynamic process.
\section{Finite dimensional dynamics}
In this section, we describe a method of finding solutions for scalar evolutionary equations by means of finite dimensional dynamics (see also~\cite{KrugLych,LychLych,AKL}). The basic idea of this method is to find finite-dimensional subspaces in an infinite-dimensional space of solutions of evolutionary equations.

First of all, let us recall some basic ideas of geometric theory of ODEs~\cite{KrVin,KLR}. Let $\pi\colon\mathbb{R}^{2}\to\mathbb{R}$ be a trivial bundle and let $J^{k}(\pi)$ with canonical coordinates $(x,y_{0},\ldots,y_{k})$ be a space of $k$-jets of sections of $\pi$. Then, any ODE $F\left(x,y,y^{\prime},\ldots,y^{(k)}\right)=0$ can be understood as a submanifold
\begin{equation}
\label{ODEdyn}
\mathcal{E}=\left\{F(x,y_{0},y_{1},\ldots,y_{k})=0\right\}\subset J^{k}(\pi).
\end{equation}
The Cartan distribution $\mathcal{C}$ on $J^{k}(\pi)$ is generated by the Cartan forms
\begin{equation*}
\omega_{j}=dy_{j}-y_{j+1}dx,\quad j=\overline{0,k-1},
\end{equation*}
or, equivalently, $\mathcal{C}=\langle\partial_{y_{k}},\mathcal{D}\rangle$, where
\begin{equation*}
\mathcal{D}=\partial_{x}+y_{1}\partial_{y_{0}}+\ldots+y_{k}\partial_{y_{k-1}}.
\end{equation*}
We will assume that $\mathcal{E}$ is a smooth submanifold and that at any point $\theta_{k}\in\mathcal{E}$ the Cartan subspace $\mathcal{C}_{\theta_{k}}$ is transversal to the tangent space $T_{\theta_{k}}\mathcal{E}$, which means that the following conditions hold:
\begin{equation*}
\frac{\partial F}{\partial y_{k}}\ne0,\quad \mathcal{D}(F)\ne0 \text{ on }\mathcal{E}.
\end{equation*}
The last implies that restriction $\mathcal{C}(\mathcal{E})$ of the Cartan distribution $\mathcal{C}$ on $\mathcal{E}$ is a one-dimensional distribution generated by a vector field
\begin{equation*}
X_{F}=\frac{\partial F}{\partial y_{k}}\mathcal{D}-\mathcal{D}(F)\frac{\partial}{\partial y_{k}}.
\end{equation*}

A vector field $X\in D(\mathcal{E})$ is called \textit{infinitesimal symmetry} of $\mathcal{E}$ if it preserves the distribution $\mathcal{C}(\mathcal{E})$, i.e. $[X,\mathcal{C}(\mathcal{E})]\subset\mathcal{C}(\mathcal{E})$. Such vector fields form a Lie algebra $\mathrm{Sym}(\mathcal{E})$ which is decomposed into a direct sum
\begin{equation*}
\mathrm{Sym}(\mathcal{E})=\mathrm{Shuff}(\mathcal{C}(\mathcal{E}))\oplus\mathcal{C}(\mathcal{E}),
\end{equation*}
where $\mathrm{Shuff}(\mathcal{E})$ is a Lie algebra of \textit{shuffle symmetries} which consists of vertical with respect to projection $\pi_{k}\colon J^{k}(\pi)\to\mathbb{R}$ vector fields on $\mathcal{E}$ preserving $\mathcal{C}(\mathcal{E})$. We will mainly be interested in ODEs resolved with respect to the higher derivative, i. e. $F=y_{k}-f(x,y_{0},\ldots,y_{k-1})$.
\begin{theorem}
Shuffle symmetries of $\mathcal{E}$ have the following form
\begin{equation*}
S_{\phi}=\sum\limits_{j=0}^{k}\mathcal{D}^{j}(\phi)\frac{\partial}{\partial y_{j}},
\end{equation*}
where $\phi\in C^{\infty}(\mathcal{E})$ is a generating function for the symmetry $S_{\phi}$ satisfying the Lie equation
\begin{equation}
\label{lie}
S_{\phi}(F)=0\,\mod\langle F,\mathcal{D}F,\ldots\rangle.
\end{equation}
\end{theorem}

Let $y=h(x)$ be a solution of equation $\mathcal{E}$ and let
\begin{equation*}
\Gamma=\left\{y_{0}=h(x),\, y_{1}=h^{\prime}(x),\ldots,y_{k}=h^{(k)}(x)\right\}\subset{\mathcal{E}}
\end{equation*}
be its prolongation to $J^{k}(\pi)$. Let $\Phi_{t}$ be a flow of the vector field $S_{\phi}$. Then, for small $t$, a curve $\Gamma_{t}=\Phi_{t}(\Gamma)\subset\mathcal{E}$ is a k-jet of $h_{t}(x)=\left(\Phi_{t}^{*}\right)^{-1}(h(x))$:
\begin{equation*}
\Gamma_{t}=\left\{y_{0}=h_{t}(x),\, y_{1}=h^{\prime}_{t}(x),\ldots,y_{k}=h^{(k)}_{t}(x)\right\}\subset{\mathcal{E}}.
\end{equation*}
Since $\dot y_{j}=\mathcal{D}^{j}(\phi)$ for $j=\overline{1,k}$, the function $u(t,x)=h_{t}(x)$ satisfies the following evolutionary equation:
\begin{equation}
\label{evol}
\frac{\partial u}{\partial t}=\phi\left(x,u,\frac{\partial u}{\partial x},\ldots,\frac{\partial^{k}u}{\partial x^{k}}\right).
\end{equation}
In other words, solutions for evolutionary equation~(\ref{evol}) can be obtained from solutions of the ODE $\mathcal{E}$ with the symmetry $S_{\phi}$ as $u(t,x)=\left(\Phi_{t}^{*}\right)^{-1}(h(x))$, and the ODE $\mathcal{E}$ is called \textit{finite dimensional dynamics} for evolutionary equation~(\ref{evol}).

The following theorem~\cite{LychLych,AKL} is used to find finite-dimensional dynamics.
\begin{theorem}
Equation~(\ref{ODEdyn}) is a finite-dimensional dynamics for evolutionary equation~(\ref{evol}) if and only if
\begin{equation*}
[\phi,F]=aF+b\mathcal{D}(F),
\end{equation*}
where $a$ and $b$ are some functions and $[\phi,F]$ is the Poisson-Lie bracket between functions $\phi$ and $F$ of the form
\begin{equation*}
[\phi,F]=\sum\limits_{j=0}^{k}\left(\frac{\partial\phi}{\partial y_{j}}\mathcal{D}^{j}(F)-\frac{\partial F}{\partial y_{j}}\mathcal{D}^{j}(\phi)\right).
\end{equation*}
\end{theorem}
In other words,
\begin{equation}
\label{PLbr}
[\phi, F]=0\,\,\mod\langle F,\,\mathcal{D}F,\ldots\rangle.
\end{equation}

\begin{definition}{\cite{AKLdan}}
Dynamics~(\ref{ODEdyn}) is said to be an \textit{attractor} for the solution $u(t,x)$ of evolutionary equation~(\ref{evol}) if
\begin{equation*}
\lim\limits_{t\to\infty}F[u]=0,
\end{equation*}
where $F[u]=F(x,u,u_{x},...,u_{x...x})$.
\end{definition}
If the dynamics found turns out to be an attractor, then one can conclude that any solution of~(\ref{evol}) behaves like that obtained by finite-dimensional dynamics method and the corresponding finite-dimensional subspaces in solution space of~(\ref{evol}) are stable.

Define functions~\cite{AKLdan}
\begin{equation*}
\psi_{1}=\frac{\partial\phi}{\partial y_{0}}+a,\quad\psi_{2}=\frac{\partial\phi}{\partial y_{1}}+b,\quad\psi_{3}=\frac{\partial\phi}{\partial y_{2}}.
\end{equation*}
The following theorem~\cite{AKLdan} provides conditions under which dynamics~(\ref{ODEdyn}) is an attractor for~(\ref{evol}).
\begin{theorem}
Let $u(t,x)$ be a solution of~(\ref{evol}) and let $\psi_{j}[u]$, $j=1,..,3$, be bounded functions. Assume that
\begin{equation}
\label{attr-ineq}
\psi_{1}[u]\le c_{1}<0,\quad\psi_{3}[u]\ge c_{2}>0.
\end{equation}
Then, dynamics~(\ref{ODEdyn}) is an attractor for~(\ref{evol}).
\end{theorem}

\section{First order dynamics and exact solutions}
Now that we have recalled all necessary constructions from thermodynamics and geometry of differential equations, we can construct solutions for non-stationary filtration equations and having all the thermodynamic variables as functions of $t$ and $x$ locate the coexistence curve obtained above in the plane $(t,x)$.

First of all, let us assume that a thermodynamic state of the gas is given by $L$ and the gas is involved in a thermodynamic process $l\subset L$ and let $\rho$ be a parameter on $l$. This means that all the thermodynamic variables are expressed in terms of $\rho$:
\begin{equation}
\label{proc}
p=p(\rho),\quad\mu=\mu(\rho),\quad k=k(\rho).
\end{equation}
\begin{theorem}
Equations~(\ref{darcy})-(\ref{cont}) for the gas $L$ and process $l\subset L$ are equivalent to the equation
\begin{equation}
\label{evolfiltr}
\rho_{t}=(Q(\rho))_{xx},
\end{equation}
where
\begin{equation*}
Q(\rho)=\int\frac{k(\rho)\rho p^{\prime}(\rho)}{\mu(\rho)}d\rho.
\end{equation*}
\end{theorem}

Let us rewrite equation~(\ref{evolfiltr}) in the form
\begin{equation}
\label{evol1}
\rho_{t}=A(\rho)\rho_{xx}+A^{\prime}(\rho)(\rho_{x})^{2},
\end{equation}
where $A(\rho)=Q^{\prime}(\rho)$. We are looking for the first order dynamics for~(\ref{evol1}) in the form~\cite{AKL}:
\begin{equation}
\label{dyn1111}
F(y_{0},y_{1})=y_{1}-f(y_{0})=0,
\end{equation}
which has symmetry
\begin{equation*}
\phi=A(y_{0})y_{2}+A^{\prime}(y_{0})(y_{1})^{2}.
\end{equation*}
The Poisson-Lie bracket~(\ref{PLbr}) therefore has the following form:
\begin{equation*}
2A^{\prime}(y_{0})f^{\prime}+fA^{\prime\prime}(y_{0})+A(y_{0})f^{\prime\prime}=0.
\end{equation*}
Its general solution is
\begin{equation*}
f(y_{0})=\frac{C_{1}y_{0}+C_{2}}{A(y_{0})},
\end{equation*}
where $C_{1}$ and $C_{2}$ are arbitrary constants.

Therefore, the first order dynamics for~(\ref{evol1}) is
\begin{equation*}
y^{\prime}=\frac{C_{1}y+C_{2}}{A(y)},
\end{equation*}
and its solution is given by
\begin{equation}
\label{soldyn}
x=\int\limits_{y^{(0)}}^{y}\frac{A(\xi)}{C_{1}\xi+C_{2}}d\xi,
\end{equation}
where $y^{(0)}$ is a constant.

Let us now get solution for~(\ref{evol1}). The symmetry $S_{\phi}$ for equation~(\ref{dyn1111}) is of the form
\begin{equation*}
S_{\phi}=\frac{C_{1}(C_{1}y_{0}+C_{2})}{A(y_{0})}\frac{\partial}{\partial y_{0}},
\end{equation*}
and its flow $\Phi_{t}\colon(x,y_{0})\mapsto(x, z(y_{0},t))$, where $z(y_{0},t)$ is found from the relation
\begin{equation*}
\int\limits_{y_{0}}^{z}\frac{A(\xi)}{C_{1}(C_{1}\xi+C_{2})}d\xi=t.
\end{equation*}
Introduce a new function $G(\xi)$ by the following way:
\begin{equation*}
G^{\prime}(\xi)=\frac{A(\xi)}{C_{1}(C_{1}\xi+C_{2})}.
\end{equation*}
Then, $G(z)-G(y_{0})=t$, from follows that $z=G^{-1}(t+G(y_{0}))$. Therefore by means of $\Phi_{t}^{-1}$ solution~(\ref{soldyn}) is transformed into solution of~(\ref{evol1}) given implicitly:
\begin{equation}
\label{sol1}
x=\int\limits_{y^{(0)}}^{G^{-1}(-t+G(\rho))}\frac{A(\xi)}{C_{1}\xi+C_{2}}d\xi.
\end{equation}
Let us consider~(\ref{sol1}) in more details. Introduce another function $H(\xi)$ by the following way:
\begin{equation*}
H^{\prime}(\xi)=\frac{A(\xi)}{C_{1}\xi+C_{2}}.
\end{equation*}
Note that $G(\xi)$ and $H(\xi)$ are related by $H(\xi)=C_{1}G(\xi)+\alpha_{1}$, where $\alpha_{1}$ is a constant. Therefore~(\ref{sol1}) is transformed as
\begin{equation*}\begin{split}
x&=H(G^{-1}(-t+G(\rho)))-H(y^{(0)})=C_{1}G(G^{-1}(-t+G(\rho)))+\alpha_{1}-H(y^{(0)})={}\\&=C_{1}(-t+G(\rho))-\alpha_{0},
\end{split}
\end{equation*}
where $\alpha_{0}=-\alpha_{1}+H(y^{(0)})$ is a constant.

Summarizing above discussion, we conclude that the following theorem is valid:
\begin{theorem}
Solution for non-stationary filtration equation~(\ref{evolfiltr}) is given as
\begin{equation}
\label{solfinal}
\rho(t,x)=G^{-1}\left(\frac{x+\alpha_{0}}{C_{1}}+t\right).
\end{equation}
where
\begin{equation*}
G(\rho)=\int\frac{Q^{\prime}(\rho)}{C_{1}(C_{1}\rho+C_{2})}d\rho,
\end{equation*}
and $C_{1}$, $C_{2}$, $\alpha_{0}$ are constants.
\end{theorem}

By means of state equations~(\ref{legvdw}) and~(\ref{proc}) one can get all the thermodynamic variables as functions of $(t,x)$.
\subsection{Isentropic processes}
From now and on, we will assume $\mu$ and $k$ to be constants for simplicity.

\begin{figure}[h!]
\centering
\includegraphics[scale=.3]{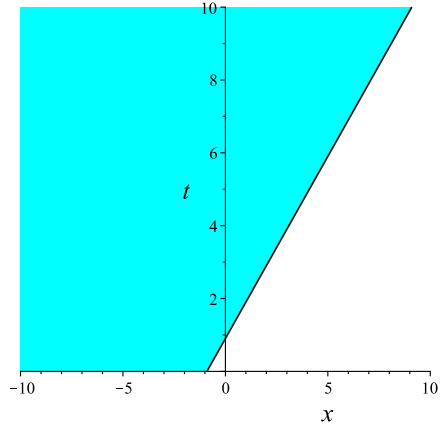}
\caption{The distribution of phases for van der Waals gases. Coloured domain is the condensation of the gas, while white one is the gas phase.}
\label{phase}       
\end{figure}

In case of isentropic processes $\sigma=R\sigma_{0}$ the pressure is expressed in terms of density in the following way:
\begin{equation*}
p(\rho)=8\exp\left(\frac{3\sigma_{0}}{4n}\right)(3\rho^{-1}-1)^{-1-2/n}-3\rho^{2}.
\end{equation*}
The function $A(\rho)=Q^{\prime}(\rho)$ is of the form
\begin{equation*}
Q^{\prime}(\rho)=\frac{k}{\mu}\left(-6\rho^{2}+24\exp\left(\frac{3\sigma_{0}}{4n}\right)\left(1+\frac{2}{n}\right)\rho^{2/n+1}(3-\rho)^{-2-2/n}\right).
\end{equation*}
The conditions for invertibility of $G(\rho)$ are given by the following theorem~\cite{LychGSA,LRljm}:
\begin{theorem}
Function $G(\rho)$ is invertible if the specific entropy constant $\sigma_{0}$ satisfies the following inequality:
\begin{equation*}
\exp\left(\frac{3\sigma_{0}}{4n}\right)>\frac{1}{4\nu}(1+\nu)^{1+\nu}(2-\nu)^{2-\nu},
\end{equation*}
where $\nu=1+2/n$.
\end{theorem}
The distribution of phases is shown in figure~\ref{phase}.
\subsection{Isenthalpic processes}
\begin{figure}[h!]
\centering
\includegraphics[scale=.4]{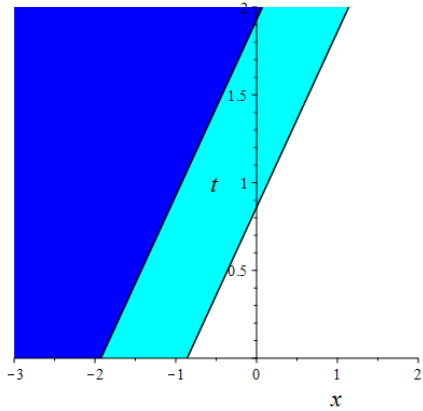}
\caption{The distribution of phases for van der Waals gases. Blue domain is the liquid phase, white domain is the gas phase, domain between is the condensation}
\label{phase1}       
\end{figure}

For isenthalpic processes $\eta=\eta_{0}$ the pressure $p(\rho)$ has the following form:
\begin{equation*}
p(\rho)=\frac{3\rho(n\rho^2+(6-3n)\rho+2\eta_{0})}{6+3n-\rho n},
\end{equation*}
and therefore
\begin{equation*}
Q^{\prime}(\rho)=\frac{6k\rho}{\mu(6+3n-\rho n)^{2}}\left(3\eta_{0}(n+2)+\rho(6-3n)(6+3n)+6n\rho^{2}(n+1)-n^{2}\rho^{3}\right).
\end{equation*}
\begin{theorem}
Function $G(\rho)$ is invertible if the specific enthalpy constant $\eta_{0}$ satisfies the following inequality:
\begin{equation*}
\eta_{0}>\frac{2(n-2)^{2}(2n+5)}{3n(n+2)}.
\end{equation*}
\end{theorem}
The distribution of phases for isenthalpic processes is shown in figure~\ref{phase1}.
\section{Second order dynamics and exact solutions for ideal gases}
In this section, we construct second order dynamics of equation~(\ref{evolfiltr}) for thermodynamic state given by ideal gas state equations:
\begin{equation*}
p=\rho RT,\quad e=\frac{n}{2}RT,\quad \sigma=R\ln\left(\frac{T^{n/2}}{\rho}\right).
\end{equation*}
We will assume that the function $A(\rho)=Q^{\prime}(\rho)$ is of the form:
\begin{equation}
\label{formA}
A(\rho)=q\rho^{\alpha},
\end{equation}
where $q>0$ and $\alpha$ are constants. One can show that at least for two processes we consider here, isenthalpic and isentropic, in case of constant viscosity $\mu$ and permeability $k$, condition~(\ref{formA}) holds. Indeed, if the level of the specific enthalpy $\eta(t,x)=\eta_{0}$ is given, one has
\begin{equation}
\label{formAenth}
A(\rho)=\frac{2\eta_{0}k\rho}{\mu(n+2)}.
\end{equation}
And if the specific entropy level $\sigma(t,x)=\sigma_{0}$ is given, we get
\begin{equation}
\label{formAent}
A(\rho)=\frac{Rk}{\mu}\left(\frac{2}{n}+1\right)\exp\left(\frac{2\sigma_{0}}{Rn}\right)\rho^{2/n+1}.
\end{equation}

We will look for the second order dynamics in the form
\begin{equation}
\label{formDyn}
F(y_{0},y_{1},y_{2})=y_{2}-y_{0}^{\beta}B(y_{1}).
\end{equation}
Such a choice is justified by form~(\ref{formA}) for $A(y_{0})$. In~\cite{AKL}, second order dynamics were found in the form $F(y_{0},y_{1},y_{2})=y_{2}-g_{1}(y_{0})y_{1}-g_{2}(y_{0})$. Such dynamics do exist but only if $A(y_{0})$ is a polynomial of degree no greater than two, which is valid for~(\ref{formAenth}) but not for~(\ref{formAent}).

The function $B(y_{1})$ is to be defined, as well as constant $\beta$. As in the previous case, one has to resolve equation~(\ref{PLbr}) with respect to $B(y_{1})$. The following theorem is a result of straightforward computations.
\begin{theorem}
Second order dynamics in the form~(\ref{formDyn}) exist if
\begin{equation*}
\beta=-1,\quad B(y_{1})=b_{1}y_{1}^{2},
\end{equation*}
where $b_{1}$ is a constant equal to either $-\alpha$, or $-\alpha+1$, or $-\alpha/2+1$.

In case of $b_{1}=-\alpha$ the corresponding dynamics is trivial.
\end{theorem}

Let us consider the last two cases.
\subsection{Case $b_{1}=-\alpha/2+1$}
Here, the second order dynamics has the following form
\begin{equation}
\label{dyn1}
y^{\prime\prime}=\left(1-\frac{\alpha}{2}\right)\frac{(y^{\prime})^{2}}{y}.
\end{equation}
Equation~(\ref{dyn1}) has two commuting symmetries
\begin{equation*}
\phi_{1}=y_{1},\quad\phi_{2}=qy_{0}^{\alpha-1}y_{1}^{2}\left(1+\frac{\alpha}{2}\right).
\end{equation*}
The first one is a translation along $x$ axis, and the second one is the right-hand side of~(\ref{evol1}). The corresponding vector fields
\begin{eqnarray*}
S_{1}&=&y_{1}\frac{\partial}{\partial y_{0}}+\frac{(2-\alpha)y_{1}^{2}}{2y_{0}}\frac{\partial}{\partial y_{1}},\\
S_{2}&=&q\left(1+\frac{\alpha}{2}\right)y_{0}^{\alpha-2}y_{1}^{2}\left(y_{0}\frac{\partial}{\partial y_{0}}+y_{1}\frac{\partial}{\partial y_{1}}\right)
\end{eqnarray*}
are linearly independent and therefore the Lie-Bianchi theorem~\cite{KrVin,KLR} can be applied to integrate~(\ref{dyn1}).

The restriction $\mathcal{C}(\mathcal{E})$ of the Cartan distribution on~(\ref{dyn1}) is given by differential 1-forms
\begin{equation*}
\omega_{1}=dy_{0}-y_{1}dx,\quad\omega_{2}=dy_{1}-\frac{(2-\alpha)y_{1}^{2}}{2y_{0}}dx.
\end{equation*}
Let us choose another basis $(\varkappa_{1},\varkappa_{2})$ in $\mathcal{C}(\mathcal{E})$ by the following way:
\begin{equation*}
\begin{pmatrix}
\varkappa_{1}\\\varkappa_{2}
\end{pmatrix}
=W^{-1}
\begin{pmatrix}
\omega_{1}\\\omega_{2}
\end{pmatrix}
,
\end{equation*}
where
\begin{equation*}
W=\begin{pmatrix}
\omega_{1}(S_{1}) & \omega_{1}(S_{2})\\
\omega_{2}(S_{1}) & \omega_{2}(S_{2})
\end{pmatrix}
.
\end{equation*}
According to the Lie-Bianchi theorem, since the Lie algebra $\langle S_{1},S_{2}\rangle$ is commutative, the new forms $(\varkappa_{1},\varkappa_{2})$ are closed and therefore locally exact, i.e. locally $\varkappa_{i}=dJ_{i}$, where $J_{i}$ are functions on $\mathcal{E}$ and solution of~(\ref{dyn1}) is given (in general, implicitly) by relations $J_{i}(x,y_{0},y_{1})=C_{i}$ for some constants $C_{i}$.

\begin{figure}[h!]
\centering
\includegraphics[scale=.3]{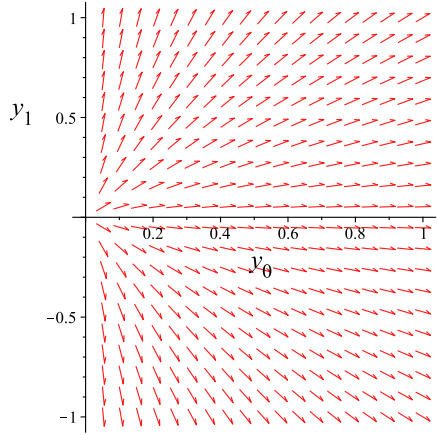}
\caption{Vector field $S_{2}$}
\label{field}       
\end{figure}

In our case, the matrix $W^{-1}$ is of the form
\begin{equation*}
W^{-1}=\frac{2}{q\alpha(\alpha+2)y_{1}^{3}}
\begin{pmatrix}
q(\alpha+2)y_{1}^{2} & -qy_{0}y_{1}(\alpha+2)\\
(\alpha-2)y_{1}y_{0}^{-\alpha+1} & 2y_{0}^{-\alpha+2}
\end{pmatrix}
,
\end{equation*}
and forms $(\varkappa_{1},\varkappa_{2})$ are
\begin{eqnarray*}
\varkappa_{1}&=&-dx+\frac{2}{y_{1}\alpha}dy_{0}-\frac{2y_{0}}{y_{1}^{2}\alpha}dy_{1},\\
\varkappa_{2}&=&\frac{2(\alpha-2)y_{0}^{-\alpha+1}}{y_{1}^{2}q\alpha(\alpha+2)}dy_{0}+\frac{4y_{0}^{-\alpha+2}}{y_{1}^{3}q\alpha(\alpha+2)}dy_{1},
\end{eqnarray*}
and the corresponding integrals are
\begin{equation}
\label{intgr}
-x+\frac{2y_{0}}{y_{1}\alpha}=C_{1},\quad \frac{2y_{0}^{-\alpha+2}}{q\alpha(\alpha+2)y_{1}^{2}}=C_{2}.
\end{equation}
Eliminating $y_{1}$ from~(\ref{intgr}) we get solution of~(\ref{dyn1}):
\begin{equation*}
y(x)=\left(\frac{2C_{2}q(\alpha+2)}{\alpha(C_{1}+x)^{2}}\right)^{-1/\alpha}.
\end{equation*}
Solution of~(\ref{evol1}) is obtained by shifting~(\ref{intgr}) along the flow of $S_{2}$. Trajectories of $S_{2}$ are shown in figure~\ref{field}.

\begin{figure}[h!]
\centering
\includegraphics[scale=.3]{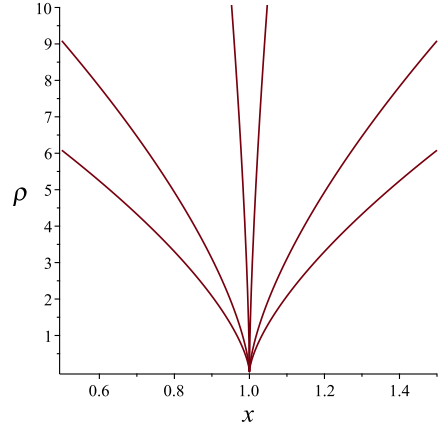}
\caption{The graph of the density.}
\label{bl-up}       
\end{figure}

The flow of the vector field $S_{2}$ has the following form
\begin{equation*}
\Phi_{t}\colon(x,y_{0},y_{1})\mapsto\left(x,y_{0}\Psi(y_{0},y_{1},t),y_{1}\Psi(y_{0},y_{1},t)\right),
\end{equation*}
where
\begin{equation*}
\Psi(y_{0},y_{1},t)=\left(1-\left(1+\frac{\alpha}{2}\right)qt\alpha y_{1}^{2}y_{0}^{\alpha-2}\right)^{-1/\alpha}.
\end{equation*}
Applying transformation $\Phi_{t}^{-1}$ to~(\ref{intgr}) and eliminating $y_{1}$, we get solution of~(\ref{evol1}):
\begin{equation}
\label{solution}
\rho(t,x)=\left(\frac{\alpha(x+C_{1})^{2}}{2q(\alpha+2)(C_{2}-t)}\right)^{1/\alpha}.
\end{equation}
For isentropic flows, $\alpha=2/n+1$ and since $n\ge 3$, the constant $\alpha$ is positive. For isenthalpic processes, $\alpha=1$. In both cases, one can observe a blow-up effect in solution. Namely, solution~(\ref{solution}) becomes infinite for all $x$ in finite time defined by constant $C_{2}$. In figure~\ref{bl-up}, the constant $C_{2}=1$, the distribution of the density is given for time moments $t=0.5$, $t=0.85$ and $t=0.999$ and it odes not exist for $t>1$.

\begin{figure}[h!]
\centering
\includegraphics[scale=.3]{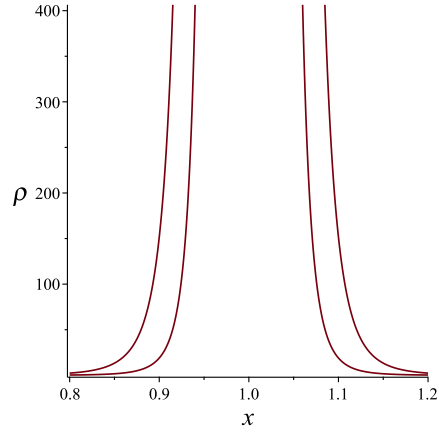}
\caption{The graph of density for flows with $\alpha=-1/3$.}
\label{figsol}       
\end{figure}

If functions $\mu(\rho)$ and $k(\rho)$ are such that $A(\rho)=q\rho^{\alpha}$, where $\alpha$ is negative, solution~(\ref{solution}) has a singularity at the point $x=-C_{1}$. The graph of the solution is given in figure~\ref{figsol} for time moments $t=5$ and $t=9$.

\subsection{Case $b_{1}=-\alpha+1$}
In this case, the second order dynamics is
\begin{equation*}
y^{\prime\prime}=(1-\alpha)\frac{(y^{\prime})^{2}}{y}.
\end{equation*}
Symmetries $S_{1}$ and $S_{2}$ are
\begin{equation*}
S_{1}=y_{1}\frac{\partial}{\partial y_{0}}+\frac{(1-\alpha)y_{1}^{2}}{y_{0}}\frac{\partial}{\partial y_{1}},\quad S_{2}=qy_{0}^{\alpha-1}y_{1}S_{1}.
\end{equation*}
Since vector fields $S_{1}$ and $S_{2}$ are linearly dependent, one cannot apply the Lie-Bianchi theorem but one still can write a solution
\begin{equation}
\label{odesol2}
y(x)=(\alpha(C_{1}x+C_{2}))^{1/\alpha}.
\end{equation}
By shifting~(\ref{odesol2}) along the trajectories of $S_{2}$ as it was done in the previous case, we get solution of the form
\begin{equation}
\label{trwave}
\rho(t,x)=\left(C_{1}^{2}\alpha qt+C_{1}\alpha x+C_{2}\alpha\right)^{1/\alpha}.
\end{equation}
Solution~(\ref{trwave}) is a travelling wave, which coincides with~(\ref{solfinal}) by choosing constants $C_{1}$ and $C_{2}$.
\subsection{Attractors}
\begin{figure}[h]

\begin{minipage}[h]{0.32\linewidth}
\center{\includegraphics[scale=0.22]{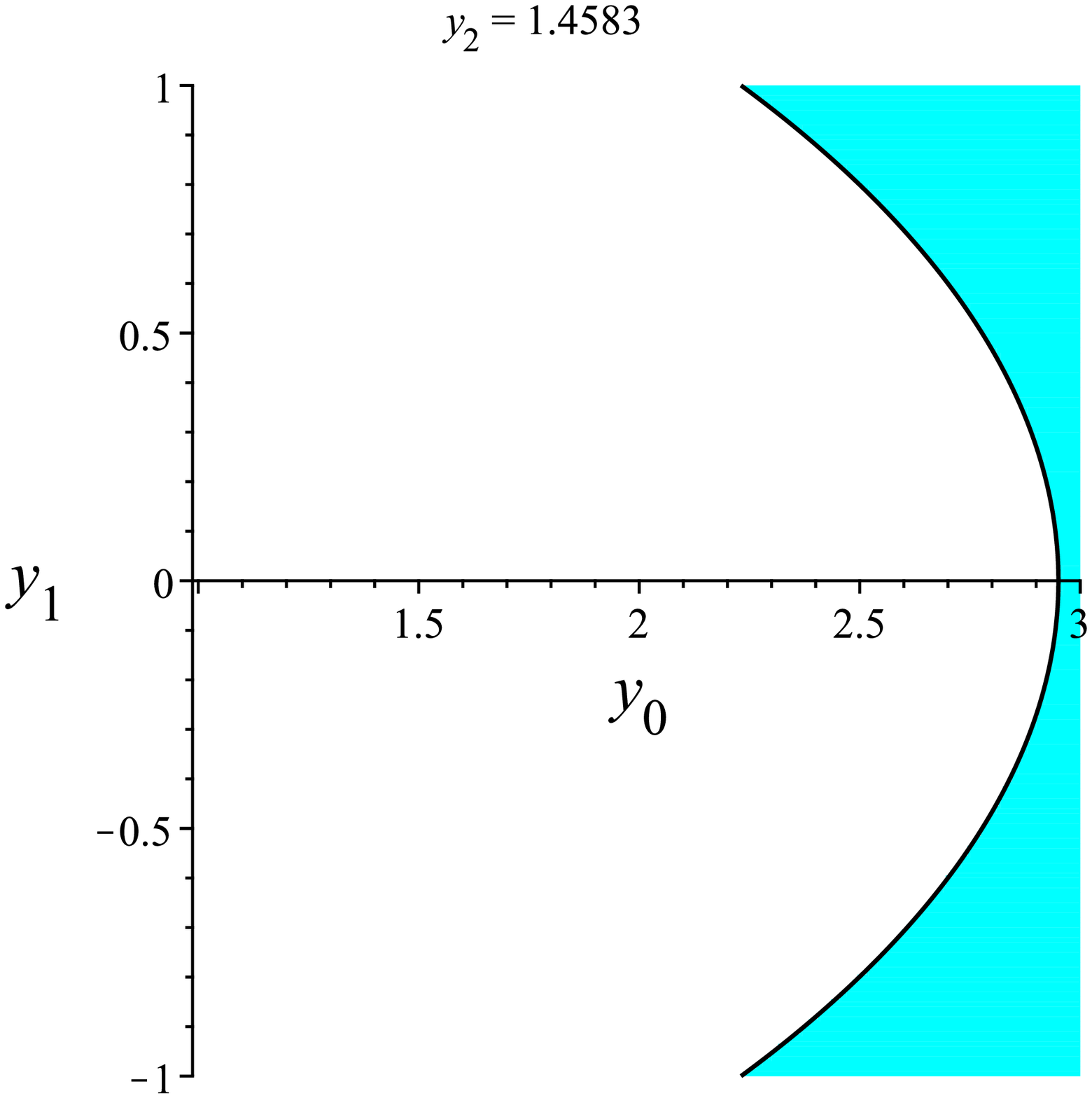}}
\end{minipage}
\hfill
\begin{minipage}[h]{0.32\linewidth}
\center{\includegraphics[scale=0.22]{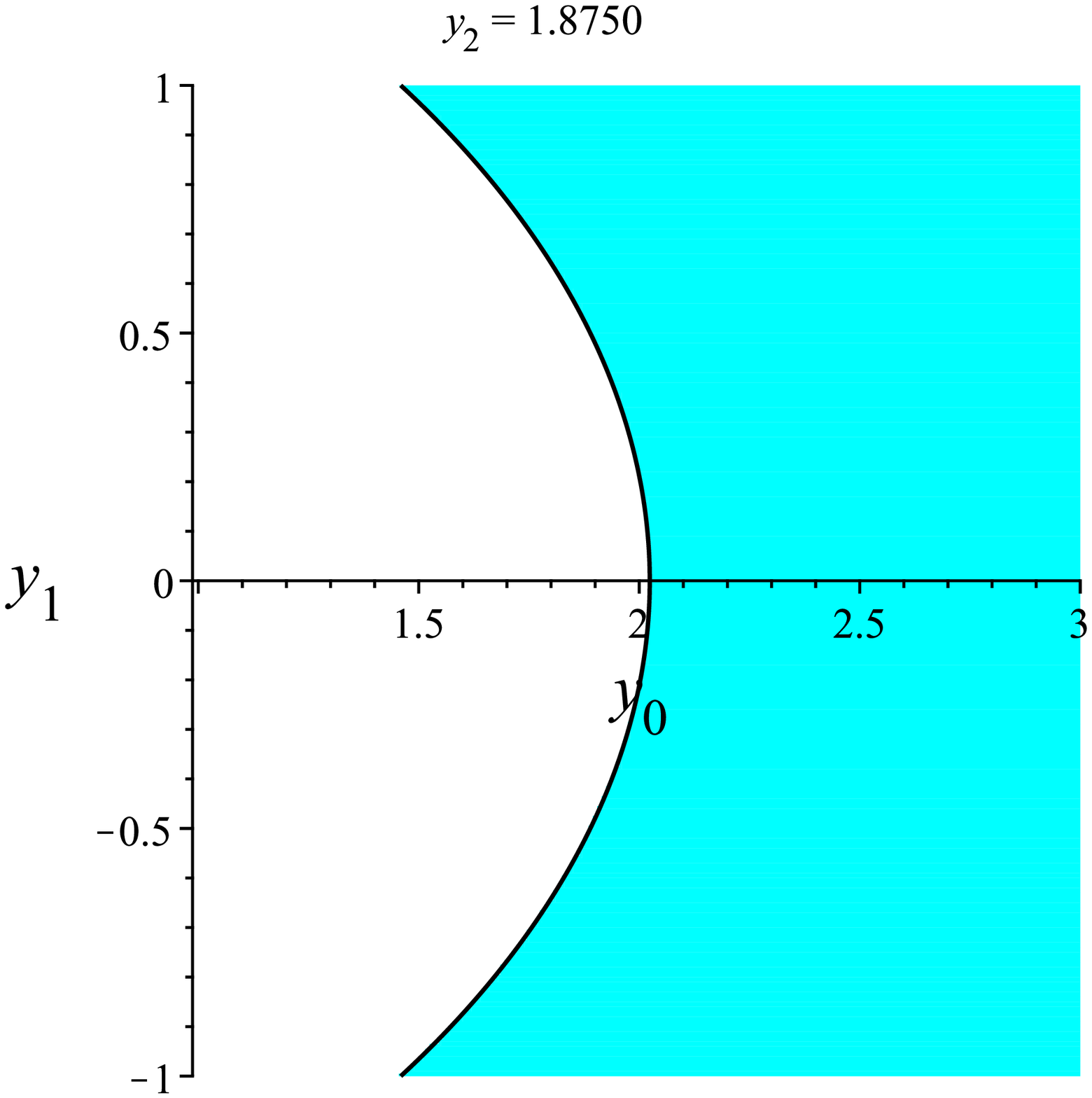}}
\end{minipage}
\hfill
\begin{minipage}[h]{0.32\linewidth}
\center{\includegraphics[scale=0.22]{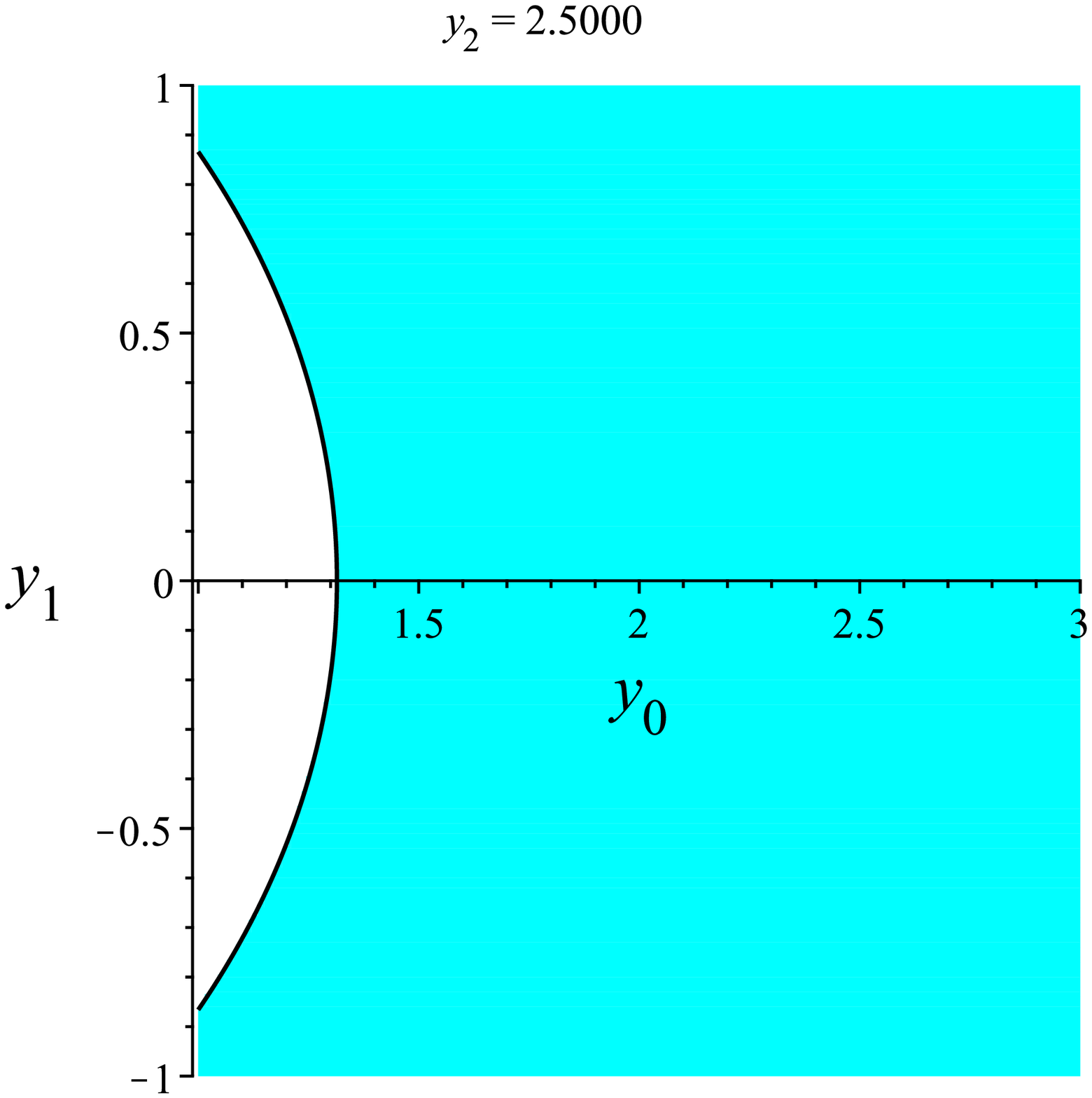}}
\end{minipage}
\caption{Attractor domains (coloured) for $q=1$, $n=3$, $c_{1}=-1$, $c_{2}=1$}
\label{attrdom}
\end{figure}

Functions $a$ and $b$ in~(\ref{PLbr}) are of the form
\begin{equation*}
a=\frac{q(\alpha+b_{1})y_{0}^{\alpha-2}\left(\left(\alpha^{2}-\alpha(b_{1}+3)+2\right)y_{1}^{4}+4y_{2}y_{1}^{2}y_{0}(\alpha-1)+2y_{2}^{2}y_{0}^{2}\right)}{b_{1}y_{1}^{2}-y_{2}y_{0}},\quad b=0,
\end{equation*}
where $b_{1}$ is either $-\alpha+1$ or $-\alpha/2+1$. Therefore, functions $\psi_{j}$ are
\begin{eqnarray*}
\psi_{1}&=&q\alpha y_{0}^{\alpha-2}\left(y_{2}y_{0}+(\alpha-1)y_{1}^{2}\right)+a,\\
\psi_{2}&=&2q\alpha y_{1}y_{0}^{\alpha-1},\quad \psi_{3}=qy_{0}^{\alpha}.
\end{eqnarray*}
For $b_{1}=-\alpha+1$ and for isentropic processes where $\alpha=2/n+1$ inequalities~(\ref{attr-ineq}) take the form
\begin{equation}
\label{ineqadiab}
-qn^{-2}(n-2)\left(y_{2}y_{0}n+2y_{1}^{2}\right)y_{0}^{2/n-1}\le c_{1}<0,\quad qy_{0}^{2/n+1}\ge c_{2}>0.
\end{equation}
Inequalities~(\ref{ineqadiab}) for given $c_{1}$ and $c_{2}$ define domains in the jet space $J^{2}(\pi)$ where dynamics~(\ref{formDyn}) is an attractor for equation~(\ref{evolfiltr}). Sections of these domains for various values $y_{2}$ are shown in figure~\ref{attrdom}.

\section*{Acknowledgements}
This work was partially supported by the Russian Foundation for Basic Research (project 18-29-10013).

\end{document}